\documentclass[twocolumn,showpacs,preprintnumbers,amsmath,amssymb]{revtex4}
\usepackage{graphicx}% Include figure files
\usepackage{dcolumn}% Align table columns on decimal point
\usepackage{bm}% bold math

\catcode`\^^?=9 % ignore

\def\be{\begin{equation}}
\def\ee{\end{equation}}
\def\ba{\begin{eqnarray}}
\def\ea{\end{eqnarray}}
\def\nn{\nonumber}
\def\ban{\begin{eqnarray*}}
\def\ean{\end{eqnarray*}}
\def\mref#1{Eq.(\ref{eq:#1})}
\def\nref#1{(\ref{eq:#1})}
\def\mlab#1{\label{eq:#1}}

\begin{document}

%\preprint{APS/123-QED}

\title{Joint Probabilities Reproducing Three
 EPR Experiments On Two Qubits }

\author{S.M.~Roy}
\altaffiliation{On leave of absence from
Department of Theoretical Physics,
Tata Institute of Fundamental Research,
Homi Bhabha Road,
Mumbai 400 005,
India.}
\affiliation{ Department of Computer Science, The University of York,
Heslington, York YO10 5DD, United Kingdom.}

\author{D.~Atkinson}
\affiliation{  Centre for Theoretical Physics, Nijenborgh 4,
9747 AG Groningen, The Netherlands.}

\author{G.~Auberson}
\affiliation{LPTA, UMR 5207 - IN2P3-CNRS, Universit\'e Montpellier II,
Montpellier, France.}

\author{G.~Mahoux}
\affiliation{CEA/Saclay, Service de Physique Th\'eorique, F-91191 Gif-sur-Yvette
Cedex, France.}

 \author{ V.~Singh}
\affiliation{Department of Theoretical Physics,
Tata Institute of Fundamental Research,
Homi Bhabha Road,
Mumbai 400 005,
India.}

\date{\today}% It is always \today, today,

\begin{abstract}
An eight parameter family of the most general nonnegative quadruple
probabilities is constructed for EPR-Bohm-Aharonov experiments when
only 3 pairs of analyser settings are used. It is a sim\-ult\-aneous
representation of 3 {\em Bohr-incompatible} experimental
configurations valid for arbitrary quantum states.
\end{abstract}

\maketitle

\noindent \textbf{Probabilities and correlations in EPR
experiments}. The EPR-Bohm-Aharonov \cite{EPR} system of two
correlated spin-half particles or qubits observed by spatially
separated observers has often been used as an arena in which to
probe fundamental questions about quantum theory. Typically there
are four different configurations corresponding to four different
experiments and one obtains no-go theorems on (i) the validity of
Einstein's local reality principle \cite{Bell} and (ii) the
existence of joint probabilities for noncommuting observables
\cite{Wigner},\cite{Fine},\cite{Belinsky}. We demonstrate here a
positive result concerning problem (ii), in the case of only three
EPR experiments, by obtaining explicitly the complete set of joint
probabilities of the relevant commuting and noncommuting
observables. Complementarity --- or the nonexistence of a joint
probability for noncommuting observables --- thus becomes a precise
quantitative issue: it does not hold for 3 EPR experiments, it does
hold for 4 EPR experiments. It should be stressed that Fine's earlier
construction of a particular joint probability for 4 EPR experiments
only holds for those quantum states which do not violate Bell-CHSH
inequalities. We obtain the most general joint probability (i) for
three EPR experiments for {\em arbitrary} quantum states, as well as
(ii) for four EPR experiments for those quantum states which obey
Bell-CHSH inequalities.

In EPR experiments one observer uses one of two possible analyser
orientations ${\bf n}_{A},{\bf n}_{A'}$ to measure dichotomic
variable $A$ or $A'$ on one qubit, with possible experimental values
$a,a'=\pm 1$, and the other observer uses one of two possible
analyser orientations ${\bf n}_{B},{\bf n}_{B'}$ to measure
dichotomic variable $B$ or $B'$ on the other qubit, with possible
experimental values $b,b'=\pm 1$. Each
experimental arrangement yields four probabilities corresponding
to the two possible results seen by each observer, of which only
three are independent, since the total probability must be unity.
With the short-hand notation of Fine \cite{Fine},
\ban
P(A)\equiv P(A=+),\;\;\; P(B)\equiv P(B=+),\nn
\\ P(AB)\equiv P(A=+, B=+),
\ean
the probabilities $P(A=a, B=b)$ with
$a=\pm 1,b=\pm 1$ can be expressed using probability sum rules:
\ban
P(A=+, B=-)= P(A)-P(AB),\nn \\ P(A=-, B=+)= P(B)-P(AB),\nn
\\P(A=-,B=-) =1-P(A)-P(B)+P(AB). \mlab{Bell}
\ean

Proceeding similarly for the other 3 experiments, we see that the 16
measured probabilities in the 4 EPR experiments can be expressed in
terms of 8 independent probabilities:
\ban
&& P(A),P(A'),P(B),P(B'),
\nn \\&& P(AB),P(AB'),P(A'B),P(A'B')\,.
\ean
The spin-spin
correlations are given in terms of these probabilities, e.g.
\ban
\langle AB\rangle = \sum _{a,b=\pm 1}\> ab \> P(A=a, B=b)
\nn  \\ =4P(AB)-2P(A)-2P(B)+1,
\ean
and the other 3 correlations are
similarly defined. It is known that Einstein's principle of local
reality requires the 4 correlations to obey Bell-CHSH inequalities
\cite{Bell},
\ba
|\langle AB\rangle+\langle AB'\rangle|&+& \nn  \\ &&
\hspace{-20mm}|\langle A'B\rangle-\langle A'B'\rangle|  \leq  2.   \mlab{Bell199}
\ea
Certain quantum density operators $\rho$
violate these local reality inequalities when the quantum
expectation values are substituted,
\[
\langle  AB\rangle \rightarrow \rm{tr} \rho A B,\> \;\;\;
 A= {\bf \sigma^{1}}\! .\, {\bf n}_{A},\;\;\;B = {\bf \sigma^{2}}\! .\,{\bf n}_{B} ,
\]
and three
analogous expressions for the other 3 correlations. Here we use
${\bf \sigma ^{1}}, {\bf \sigma ^{2}}$ to denote Pauli spin
operators for the two qubits.

We focus now on the question of the existence of joint probabilities
for noncommuting observables, first posed by Wigner \cite{Wigner}.
Each EPR experiment yields probabilities of eigenvalues $\pm 1$ for
one complete set of commuting observables, e.g. the probability of
values $a,b$ for $A,B$; but two different EPR experiments involve
noncommuting observables. Does there exist for every quantum state
a positive normalised joint quadruple probability distribution
$P(aa'bb')$ whose marginals reproduce the quantum probabilities of
all 4 EPR experiments, i.e. \ba && P(A=a,B=b)= P(a\>.\>b\>.\>),\nn
\\ && P(A=a,B'=b')= P(a\>.\>.\>b'),\nn
\\ && P(A'=a',B=b)= P(\>.\>a'b\>.\>),
\mlab{3probys} \ea and \ban P(A'=a',B'=b')= P(\>.\>a'\>.\>b'), \ean
where those of the indices $aa'bb'$ that have been replaced by dots
are to be summed over the values $\pm 1$ (for brevity simply $\pm$).
As there are only 8 independent probabilities in the 4 EPR
experiments, these 16 marginal conditions on the quadruple
probabilities imply 4 constraints from the single probabilities: \ba
\nn P(A) = P (+\>.\>.\>.\>),&& P(A') = P(\>.\>+\>.\>.\>), \\
P(B) = P(\>.\>.\>+\>.\>),&& P(B') = P(\>.\>.\>.\>+)\,; \mlab{1}
\ea
and four constraints from the double probabilities:
\ba
&& \hspace*{-10mm} P(AB)=P(+\>.\>+\>.\>),\;\;  P(AB')=P(+\>.\>.\>+), \nn \\
&& \hspace*{-10mm}  P(A'B)=P(\>.\>+ +\>.\>),\; P(A'B')= P (\>.\> +\> .\> +)\,.
\mlab{111}
\ea
\noindent \textbf{Bell-CHSH inequalities from probability sum
rules.} It is illuminating that, from Eqs.\nref{1}-\nref{111} we can rewrite the 4
constraints from the double probabilities as restrictions on certain positive
combinations of unobserved triple probabilities, namely
 \ba
 C(AA'BB') &=& P(+ + \>.\> -) + P(+ - -\> .\> ) \nn
 \\
&\>& + P(- + +\> .\> ) + P (- -\> .\> + ), \mlab{2} \ea with the
definition
 \ban
 &\>&\hspace{-10mm} C(AA'BB')\> \equiv P(A) + P(B') \nn
\\  &&\hspace*{-5mm}  - [ P(AB) + P(AB') -P (A'B) + P (A'B') ],
\mlab{112}
\ean
and three other equations obtained by interchanging $A$ and $A'$, or
$B$ and $B'$, or $A$ and $A'$ as well as $B$ and $B'$,  in the
arguments of $C$.  On the
right-hand side of \mref{2} one must
correspondingly interchange the first argument with the second,
the third argument with the fourth, and the first with the second as
well as the third with the fourth, respectively,   in the quadruple probabilities.
 Since the quadruple probabilities must
also obey a ninth constraint of normalisation,
\ban
1=P(\>.\>.\>.\>.\>),
\ean
 the right-hand side of \mref{2} and
therefore also the left-hand side must lie in the interval $[0,1]$.
Thus we obtain 8 inequalities on the measured  probabilities,
\ba
 &&\hspace{-10mm}
0 \leq C(AA'BB')\leq 1 \>,\;\;\; 0 \leq C(A'ABB') \leq 1 ,\nn
\\ &&\hspace{-10mm} 0 \leq  C(AA'B'B) \leq 1\>,\;\;\;0 \leq  C(A'AB'B) \leq 1 \>. \mlab{3}
\ea
These  are exactly the Bell-CHSH inequalities re-expressed in terms
 of probabilities. The 4
hyperplanes of the type \nref{2} in the space of the $P(aa'bb')$ allow
one to prove simply
 that the Bell-CHSH inequalities are necessary for the
existence of positive normalised quadruple probability
distributions.

\noindent \textbf{Construction of the most general positive
quadruple distribution fitting 4 EPR experiments.} In his elegant
work, Fine \cite{Fine} has shown that the Bell-CHSH inequalities are
both necessary and sufficient for the existence of positive quadruple
distributions. His existence proof however does not yield the
complete set of such distributions. The complete
set of the $P(aa'bb')$, but without constraints of positivity, has been given
by Atkinson \cite{Atkinson}. We shall here explicitly construct the
most general positive quadruple distribution fitting 4 EPR
experiments whenever Bell-CHSH inequalities are satisfied. Fine's
construction, nonlinear in the unobserved triple probabilities,
will be replaced by a linear construction to achieve this goal.
Since the 16 $P(aa'bb')$ have to fit 8 independent experimental
probabilities and 1 normalisation constraint, the most general
quadruple distribution will have 7 free parameters.

\noindent
{\bf Theorem}  \\  Suppose the 8 experimental probabilities
\ban &&
P(A), P(A'), P(B), P(B'), \\  &&  \hspace{15mm}P(AB), P(AB'),
P(A'B), P(A'B'),
\ean obey the Bell-CHSH inequalities \nref{3}. Then
there exist positive values of the seven free parameters,
\ban
P(aa'++)\;\; , \;\; P(++bb')
\ean
($aa'bb'$ taking the values $+$ or
$-$) in terms of which the most general nonnegative $P(aa'bb')$ can
be explicitly constructed.

%\vspace{1mm}

\noindent {\bf Remark:} Since $P(aa'++)$, $P(++bb')$ have $P(++++)$
in common they constitute 7 parameters ; further  we can choose
$P(a\>.\>++), P(\>.\>a'++)$  with a common value of $P(\>.\> .\>++)$
as three parameters and $P(++bb')$ as the remaining 4 parameters.

%\vspace{1mm}

\noindent
{\bf The proof will consist of two steps:}
\\ {\bf Step 1.}\\
Given $P(a\>.\>++),\,  P(\>.\> a'++)$ and the experimental
probabilities, the triple probabilities $P(a\>.\>bb')$ and
$P(\>.\>a'bb')$ are constructed as follows. First
\ba
\nn  P (a\>.\>+
-)&=& P (a\>.\>+\>.\> )  - P (a\>.\>++)
\\  P (a\>.\> -+) &=& P (a\>.\>.\>+) - P (a\>.\>++) \nn \\
 P (a\>.\>--)&=& P (a\>.\>.\>.\>)+ P
(a\>.\>++) \nn \\&&- P (a\>.\>+\>.\>) - P (a\>.\>.\>+)\,,
 \mlab{4}
 \ea
 and
are nonnegative if $P(a\>.\>++)$ is chosen to obey
\ba  &&\hspace{-10mm} {\rm
max}[0, P(a\>.\>+\>.\>) + P (a\>.\>.\>+) - P (a\>.\>.\>.\>) ]
\nn
\\  &&\mlab{5}     \leq
P(a\>.\>++) \leq {\rm min} [ P (a\>.\>+\>.), P (a\>.\>.\>+) ]\,.
\ea
Note that the region so defined for $P(a\>.\>++)$ is non-empty
because of probability sum rules obeyed by the experimental
probabilities. The triple probabilities $P (\>.\>a'bb')$ are
calculated in terms of $P(\>.\>a'++)$ in an analogous way:
 \ba
 \nn  P (\>.\>a'+ -)&=& P (\>.\>a'+\>.\> )  - P
(\>.\>a'++)
\\  P (\>.\>a' -+) &=& P (\>.\>a'.\>+) - P (\>.\>a'++)\nn \\
 P (\>.\>a'--)&=& P (\>.\>a'.\>.\>)+ P
(\>.\>a'++) \nn \\&& -P (\>.\>a'+\>.\>) - P (\>.\>a'.\>+)\,,
\mlab{6}
\ea
 and
these $P(\>.\>a'bb')$ are nonnegative if $P(\>.\>a'++)$ is chosen
to obey
\ba
&&\hspace{-12mm} {\rm max}[0, P(\>.\>a'+\>.\>) +
P (\>.\>a'\>.\>+) - P (\>.\>a'\>.\>.\>) ]
\nn   \\
&&\hspace{-7mm}  \mlab{7}  \leq P(\>.\>a'++) \leq {\rm min} [ P (\>.\>a'+\>.\>), P
(\>.\>a'\>.\>+) ].
\ea
Again, the region so defined for
$P(\>.\>a'++)$ is non-empty because of the probability sum rules obeyed
by the experimental probabilities.

However the sum over $a$ of $P(a\>.\>++)$ and the sum over $a'$ of
$P(\>.\>a'++)$ must be equal to the common  $P (\>.\> .\> ++)$, and
this leads to conditions for consistency of the above inequalities
on $P(a\>.\>++)$ and $P(\>.\>a'++)$. Before examining them, note
that \mref{4} and \mref{6} imply 4 equations like \mref{2}and hence
Bell-CHSH inequalities. E.g. if we add the third of \mref{4} with
$a=+$, the first of \mref{6} with $a'=+$ and the second of \mref{6}
with $a'=-$ , and utilise the meaning of the dots inside the
probabilities to simplify and regroup terms, the triple probability
$P(+\>.\>++)$ cancels, and we obtain \mref{2}. This immediately
implies that $C(AA'BB')$ must lie in the interval $[0,1]$. The other
6 Bell-CHSH inequalities follow similarly. We shall show that these
conditions are sufficient to construct positive quadruple
probabilities. No new conditions arise.

The allowed region for $P(\>.\>.\>++)$ derived from the
$P(a\>.\>++)$ inequalities \nref{5},
\ba
\nn &&{\rm max }[ 0,
P(\>.\>.\>+\>.\>) + P(\>.\>.\>.\>+) - P(\>.\>.\>.\>.\>) ,
\\  \nn && P(+\>.\>+\>.\>) +
P(+\>.\>.\>+) - P(+\>.\>.\>.\>) , \\  \nn && P(-\>.\>+\>.\>) + P
(-\>.\>.\>+) - P(-\>.\>.\>.\>) ]
\\  &\leq& P (\>.\>.\>++)\mlab{8} \\ \nn  &\leq&  {\rm min} [ P(\>.\>.\>+\>.\>),
 P(\>.\>.\>.\>+), \\   && P(+\>.\>+\>.\>) +P(-\>.\>.\>+), P(+\>.\>.\>+)
 + P(-\>.\>+\>.\>) ] , \nn
 \ea
and the allowed region for $P(\>.\>.\>++)$ derived from the
inequalities \nref{7},
\ba
\nn && {\rm max} [ 0, P(\>.\>.\>+\>.\>) +
P(\>.\>.\>.\>+) - P(\>.\>.\>.\>.\>) ,
\\  \nn && P(\>.\>++\>.\>) + P(\>.\>+\>.\>+) - P(\>.\>+\> .\>.\>) ,
\\  \nn && P(\>.\>-+\>.\>) +
P (\>.\>-\>.\>+) - P(\>.\>-\>.\>.\>) ] \\ &\leq&  P (\>.\>.\>++) \mlab{9} \\
&\leq& {\rm min} [ P(\>.\>.\>+\>.\>), P (\>.\>.\>.\>+),
 \nn
\\  \nn &&\hspace{-10mm}
P(\>.\>++\>.\>) +P(\>.\>-\>.\>+),  P(\>.\>+\>.\>+) + P(\>.\>-+\>.\>) ] \,, \nn
\ea
must
have non-empty intersection. Using $P(\>.\>.\>.\>.\>) = 1$, and
omitting inequalities guaranteed by the probability sum rules obeyed
by the experimental probabilities, we see that the regions \nref{8}
and \nref{9} intersect if and only if the following inequalities
hold:
\ban
\nn &&  {\rm max} [P(AB) + P(AB') - P (A), \\
\nn
&& P (B) - P (AB) + P (B') - P (AB') + P (A ) -1] \\
&\leq& {\rm min} [ P (A'B) + P (B') - P (A'B'),   \nn \\
&&\hspace{15mm}  P (A'B') + P( B ) - P (A'B) ], \mlab{10}
\ean
and
another set of inequalities obtained by interchanging $A$ and  $A'$:
\ban
\nn &&  {\rm max} [P(A'B) + P(A'B') - P (A'), \\
\nn
&& P (B) - P (A'B) + P (B') - P (A'B') + P (A' ) -1] \\
&\leq& {\rm min} [ P (AB) + P (B') - P (AB'),  \nn \\
&&\hspace{15mm}  P (AB') + P( B ) - P (AB) ]\,. \mlab{11}
\ean
Together these are exactly equivalent to the
8 Bell-CHSH inequalities. Hence the $P(a\>.\>bb')$ and
$P(\>.\>a'bb')$ given by the above construction, viz. \mref{4} and
\mref{6}, are nonnegative  if and only if the Bell-CHSH
inequalities hold.

 \noindent {\bf Step 2.} \\   Imitating the procedure of Step 1,
given the freely chosen 4 unobserved probabilities $P (++bb')$, and
the triple probabilities $P(a\>.\>bb')$ and $P(\>.\>a'bb')$ obtained
in Step 1, we obtain the other $P(aa'bb')$ as follows:
\ba
P (+ -
bb') &=& P (+ \>.\> bb') - P ( + + bb') \nn \\ \nn P ( - + bb') &=&
P (\>.\> + bb') - P (+ + bb') \\  P ( - - bb') &=& P (\>.\> .\> bb')
- P (\>.\> + bb')\nn
\\  && \hspace{0mm} - P (+\> .\> bb') + P
( + + bb')  .   \mlab{12}
\ea
All the $P( aa'bb')$ are nonnegative if $P (+ +
bb')$ is chosen to obey
\ba
&& \hspace{-8mm}{\rm max} [0, P(+\>.\>bb') +
P(\>.\>+bb') - P(\>.\> .\> bb') ] \nn \\ & \leq & P(++bb') \leq {\rm
min} [P( + \>.\> bb'), P(\>.\> +bb') ]. \mlab{13}
\ea
Note that this
allowed region is non-empty, since the constructed triple
probabilities obey the sum rules
\ban
  P( \>.\> .\> bb') &=& P
(\>.\>+bb' ) + P (\>.\> - bb') \nn \\ &=&  P (+\> .\> bb') + P (-\>
.\> bb')\,,
\ean
as a consequence of the sum rules that are obeyed by the input
experimental probabilities. This concludes the construction.

\noindent \textbf{Summary}. Suppose the Bell-CHSH inequalities are
obeyed. Choose 3 free parameters $P(a\>.\>++), P(\>.\>a'++)$,  with a
common value of $P(\>.\> .\>++)$, in the region given by \mref{5} and
\mref{7}. That this is possible is due to the validity of Bell-CHSH
inequalities. Calculate the triple probabilities $P(a\>.\>bb')$ and
$P(\>.\>a'bb')$, using  Eqs.\nref{4}-\nref{6}. Choose the 4 free
parameters $P (++bb')$ in the region \mref{13}. Calculate the
remaining $P(aa'bb')$ using \mref{12}. Their positivity is
guaranteed.

\noindent \textbf{Construction of the most general positive
quadruple distribution fitting three EPR experiments.} The previous
construction for 4 EPR experiments only works for those quantum
states which yield correlations obeying the Bell-CHSH inequalities.
Now suppose only 3 EPR experiments, e.g. those measuring the
probabilities given by \mref{3probys}, have been performed, but the
probabilities $ P(A'=a',B'=b')$ have not been measured. We show
below that for arbitrary quantum states we can construct the most
general positive normalised quadruple probabilities fitting all
three EPR experiments.

Of the eight experimental probabilities considered in the last
section, one unmeasured double probability, viz. $P(A'B')$, becomes
an extra free parameter. We can choose this parameter, or
equivalently the unmeasured $\langle A'B'\rangle $, to be consistent with the
Bell-CHSH inequalities \mref{Bell199},
\ban \nn &&\hspace{-5mm}
|4P(A'B)-4P(A'B')-2P(B)+2P(B')|\\&& \hspace{22mm}\leq 2-|\langle AB+AB'\rangle |.
\ean
Since the right-hand side is nonnegative, such a choice of
$P(A'B')$ is always possible. With $P(A'B')$ so chosen, we use the
steps 1 and 2 of the last section to choose the other 7 free
parameters in the regions derived there and calculate the positive
quadruple probabilities. This completes the construction of the most
general nonnegative quadruple probabilities fitting three EPR
experiments. They contain 8 free parameters.

\noindent\textbf{Conclusions.} The constructed 8 parameter set of
positive quadruple probabilities fit the observables of 3 EPR
experiments for arbitrary quantum states. Of course for those states
that violate the Bell-CHSH inequalities for 4 EPR experiments, the
parameter $P(A'B')$ in our construction cannot agree with the
predicted quantum probability for the fourth experiment.
Nevertheless what is remarkable is the existence of a joint
probability distribution whose marginals reproduce quantum
probabilities for three different complete commuting sets (CCS) of
observables $(A,B), (A,B'),(A',B)$, corresponding to three different
experimental arrangements. This is a limited breakdown of the
complementarity of noncommuting observables; it does not extend to
four EPR experiments unless the Bell inequalities are satisfied. One
may speak of ``Bohr-incompatible'' experiments that nonetheless may
all be described by the same probabilities. This result is similar
to and inspired by the ``three marginal theorem" for continuous
variables conjectured \cite{Roy} and proved recently
\cite{Auberson}. It is very different from the realization of
unsharp measurements of noncommuting observables which are not
described by projection operators but by positive operator-valued
measures (POVM) \cite{Busch},  which are a set of noncommuting
positive operators summing to the identity operator. What we have
exhibited is a simultaneous realization of probabilities of
noncommuting observables contained in three different CCS of
observables, each consisting of standard von Neumann projection
operators. It may be possible to generalize this result to
incorporate 3 different POVM's. It might be interesting to look
at two qubits with more than two analyser settings for each qubit
\cite{RoySingh}, and at three qubits, in order to understand the unfolding
lessons. It {\em would} be interesting to investigate whether the
present results can help build an extended measurement theory.

 One of us (SMR) is funded in part by EPSRC grant EP/D500354/1.


\begin{thebibliography}{99}
\bibitem{EPR}
A. Einstein, B. Podolsky and N. Rosen, Phys. Rev. {\bf 47}, 777 (1935).
\bibitem{Bell}
J.S. Bell, Physics {\bf 1}, 195 (1964).
 J.F. Clauser, M.A. Horne,
A. Shimony and R.A. Holt, Phys. Rev. Lett. {\bf 23}, 880 (1969).
\bibitem{Wigner}
E.P. Wigner, Am. J. Phys. {\bf 38}, 1005 (1970).
\bibitem{Fine}
A. Fine, Phys. Rev. Lett. {\bf 48}, 291 (1982).
\bibitem{Belinsky}
A.V. Belinsky, Physics-Upsekhi {\bf 37}, 219, 413 (1994).
\bibitem{Atkinson}
D. Atkinson, Pramana {\bf 56}, 139 (2001).
\bibitem{Roy}
S.M. Roy and V. Singh, Phys. Lett. {\bf A255}, 201 (1999).
\bibitem{Auberson}
G. Auberson, G. Mahoux, S.M. Roy and V. Singh,
 Phys. Lett. {\bf A300}, 327 (2002) and Jour. Math. Phys. {\bf 44},
2729 (2003).
\bibitem{Busch}
P. Busch, M. Grabowsky and P. J. Lahti, {\it Operational Quantum
Physics}, Springer-Verlag Berlin Heidelberg (1995).
\bibitem{RoySingh}
S. M. Roy and V. Singh, J. Phys. {\bf A11}, L167 (1978); A. Garg and
N. D. Mermin,Phys. Rev. Lett. {\bf 49}, 1220 (1982); N. Gisin, Phys.
Lett. {\bf A260}, 1 (1999).

\end{thebibliography}
\end{document}